\documentclass[useAMS,usenatbib]{mn2e}
\usepackage{epsf,graphicx,rotating,url}

\title[Coronal line variability in NGC 5548]{Strong variability of the coronal line region in NGC~5548}

\author[H. Landt et al.]{Hermine Landt$^1$\thanks{E-mail:
    hermine.landt@durham.ac.uk}\thanks{Visiting Astronomer at the
    Infrared Telescope Facility, which is operated by the University
    of Hawaii under Cooperative Agreement no. NNX-08AE38A with the
    National Aeronautics and Sp ace Administration, Science Mission
    Directorate, Planetary Astronomy Program.}, Martin J. Ward$^1$\footnotemark[2], Katrien C. Steenbrugge$^2$ and Gary J. Ferland$^{3,4}$ \\
$^1$Department of Physics, Durham University, South Road, Durham DH1 3LE, UK \\
$^2$Instituto de Astronom\'ia, Universidad Cat\'olica del Norte, Avenida Angamos 0610, 1270709 Antofagasta, Chile \\
$^3$School of Mathematics and Physics, Queen's University of Belfast, Belfast BT7 1NN, Northern Ireland, UK \\
$^4$Department of Physics and Astronomy, University of Kentucky, Lexington, KY 40506, USA}

\begin{document}

\def\la{\mathrel{\hbox{\rlap{\hbox{\lower4pt\hbox{$\sim$}}}\hbox{$<$}}}}
\def\ga{\mathrel{\hbox{\rlap{\hbox{\lower4pt\hbox{$\sim$}}}\hbox{$>$}}}}

\font\sevenrm=cmr7

\def\SII{[S~{\sevenrm II}]}
\def\SIII{[S~{\sevenrm III}]}
\def\SVIII{[S~{\sevenrm VIII}]}
\def\SIX{[S~{\sevenrm IX}]}
\def\SXI{[S~{\sevenrm XI}]}
\def\SXII{[S~{\sevenrm XII}]}

\def\SiII{[Si~{\sevenrm II}]}
\def\SiIII{[Si~{\sevenrm III}]}
\def\SiVI{[Si~{\sevenrm VI}]}
\def\SiVII{[Si~{\sevenrm VII}]}
\def\SiVIII{[Si~{\sevenrm VIII}]}
\def\SiIX{[Si~{\sevenrm IX}]}
\def\SiX{[Si~{\sevenrm X}]}
\def\SiXI{[Si~{\sevenrm XI}]}

\def\OI{[O~{\sevenrm I}]}
\def\OII{[O~{\sevenrm II}]}
\def\OIII{[O~{\sevenrm III}]}

\def\OVI{O~{\sevenrm VI}}
\def\OVII{O~{\sevenrm VII}}
\def\OVIII{O~{\sevenrm VIII}}
\def\OIX{O~{\sevenrm IX}}

\def\NeIX{Ne~{\sevenrm IX}}

\def\NVI{N~{\sevenrm VI}}
\def\NVII{N~{\sevenrm VII}}

\def\CIV{C~{\sevenrm IV}}
\def\CVI{C~{\sevenrm VI}}
\def\CVII{C~{\sevenrm VII}}

\def\FeII{[Fe~{\sevenrm II}]}
\def\FeVI{[Fe~{\sevenrm VI}]}
\def\FeVII{[Fe~{\sevenrm VII}]}
\def\FeX{[Fe~{\sevenrm X}]}

\def\FeVIIp{Fe~{\sevenrm VII}}

\def\HeII{He~{\sevenrm II}}

\def\cloudy{{\sevenrm CLOUDY}}

\date{Accepted ~~. Received ~~; in original form ~~}

\pagerange{\pageref{firstpage}--\pageref{lastpage}} \pubyear{2015}

\maketitle

\label{firstpage}

\begin{abstract}

We present the second extensive study of the coronal line variability
in an active galaxy. Our data set for the well-studied Seyfert galaxy
NGC~5548 consists of five epochs of quasi-simultaneous optical and
near-infrared spectroscopy spanning a period of about five years and
three epochs of X-ray spectroscopy overlapping in time with
it. Whereas the broad emission lines and hot dust emission varied only
moderately, the coronal lines varied strongly. However, the observed
high variability is mainly due to a flux decrease. Using the optical
\FeVII~and X-ray \OVII~emission lines we estimate that the coronal
line gas has a relatively low density of $n_e \sim 10^3$~cm$^{-3}$ and
a relatively high ionisation parameter of $\log U \sim 1$. The
resultant distance of the coronal line gas from the ionising source of
about eight light years places this region well beyond the hot inner
face of the dusty torus. These results imply that the coronal line
region is an independent entity. We find again support for the X-ray
heated wind scenario of Pier \& Voit; the increased ionising radiation
that heats the dusty torus also increases the cooling efficiency of
the coronal line gas, most likely due to a stronger adiabatic
expansion. The much stronger coronal line variability of NGC~5548
relative to that of NGC~4151 can also be explained within this
picture. NGC~5548 has much stronger coronal lines relative to the low
ionisation lines than NGC~4151 indicating a stronger wind, in which
case a stronger adiabatic expansion of the gas and so fading of the
line emission is expected.

\end{abstract}


\section{Introduction}

In addition to the broad and narrow emission lines, the spectra of
active galactic nuclei (AGN) display high-ionisation emission lines,
the so-called coronal lines, which require energies $\ga 100$~eV to be
excited. The coronal line region is believed to lie at distances from
the central ionising source intermediate between those of the broad
(BELR) and narrow emission line region (NELR) and to possibly coincide
with the hot inner face of the circumnuclear, obscuring dusty torus
\citep[as first suggested by][]{Pier95}. This assumption is based
mainly on the following: (i) the coronal lines have $3-4$ orders of
magnitude higher critical densities for collisional deexcitation than
the low-ionisation narrow emission lines and are believed to be
emitted at these densities; (ii) the coronal line profiles often but
not always have full widths at half maxima (FWHM) intermediate between
those of the broad and narrow emission lines \citep[FWHM$\sim
  500-1500$~km~s$^{-1}$; e.g.,][]{Pen84, App88, Erk97, Rod02c, Rod11};
and (iii) the emission from this region is often extended but much
less so than that from the low-ionisation NELR \citep[on scales of
  $\sim 80-150$~pc; e.g.,][]{Prieto05, Mueller06, Mueller11,
  Maz13}. Recently, \citet{Rose15} found evidence in their study of
seven AGN with strong coronal lines selected from the Sloan Digital
Sky Survey (SDSS) that the coronal line region has higher densities
than the low-ionisation NELR and lies at distances from the ionising
source similar to those estimated for the hot dust sublimation radius.

In \citet{L15}, we have presented the first extensive study of the
coronal line variability in an AGN and have started to question this
picture. Our data set for the nearby, well-known source NGC~4151
included six epochs of quasi-simultaneous optical and near-IR
spectroscopy spanning a period of $\sim 8$~years and five epochs of
X-ray spectroscopy overlapping in time with it, with the observations
in each wavelength performed with the same telescope and set-up. For
this source we found that the coronal lines varied only weakly, if at
all, and estimated a relatively low density ($n_e \sim
10^3$~cm$^{-3}$) and high ionisation parameter ($\log U \sim 1$) for
the coronal line gas. The resultant distance of the coronal line
region from the ionising continuum was $\sim 2$~light years, which is
well beyond the hot inner face of the obscuring torus of \citep[$\sim
  2$~light months;][]{Kosh14}. Based on the high ionisation parameter,
we proposed that the coronal line region is an independent entity
rather than part of a continuous gas distribution connecting the BELR
and low-ionisation NELR, possibly an X-ray heated wind as suggested by
\citet{Pier95}.

The most stringent constraints on the properties of the coronal line
emitting region could come from variability studies, in particular if
the variability of several coronal lines can be compared with each
other and with that of other AGN components such as the BELR, hot dust
emission and X-ray luminosity, as we have done in
\citet{L15}. However, mainly due to the weakness of these emission
lines and also lack of data, very few studies of this kind have been
attempted so far. \citet{Vei88} made the only systematic study of the
coronal line variability. In his sample of $\sim 20$ AGN he found firm
evidence that both the \FeVII~$\lambda 6087$ and \FeX~$\lambda 6375$
emission lines varied (during a period of a few years) for only one
source, namely, NGC~5548, which is the subject of this paper, and
tentative evidence for another seven sources (including
NGC~4151). Then, within a general optical variability campaign on the
source Mrk~110 lasting for half a year, \citet{Kolla01} reported
strong \FeX~variations. More recently, follow-up optical spectroscopy
of a handful of objects with unusually prominent coronal lines
selected from the SDSS showed that in half of them the coronal lines
strongly faded (by factors of $\sim 2-10$), making these sources
candidates for stellar tidal disruption events \citep{Kom09, Yang13}.

Here we present the second extensive study of the coronal line
variability in an AGN. Our data set for the well-studied source
NGC~5548 ($z=0.017$) consists of five epochs of quasi-simultaneous
optical and near-infrared spectroscopy spanning a period of $\sim
5$~years and three epochs of X-ray spectroscopy overlapping in time
with it. The paper is organised as follows. In Section 2, we present
the data and measurements. In Section 3, we discuss the observed
variability behaviour of the near-IR, optical and X-ray coronal lines,
for which we seek an interpretation in the context of the location of
and excitation mechanism for the coronal line emission region in
Section 4. Finally, in Section 5, we summarise our main results and
present our conclusions. Throughout this paper we have assumed
cosmological parameters $H_0 = 70$ km s$^{-1}$ Mpc$^{-1}$,
$\Omega_{\rm M}=0.3$, and $\Omega_{\Lambda}=0.7$.

\section{The data and measurements}

\subsection{Near-IR and optical spectroscopy}

\begin{table*}
\caption{\label{irlines} 
Near-IR emission line fluxes and ratios}
\begin{tabular}{lccccccccc}
\hline
Observation & \SIII~$\lambda$9531 & \SIII~$\lambda$9069 & \SIII/ & \SVIII~$\lambda$9911 & \SIII/ & \SIX~1.252$\mu$m & \SIII/ \\
Date & (erg/s/cm$^2$) & (erg/s/cm$^2$) & \SIII & (erg/s/cm$^2$) & \SVIII & (erg/s/cm$^2$) & \SIX \\
\hline
2002 Apr 23 & (2.79$\pm$0.06)e$-$14 & (1.24$\pm$0.04)e$-$14 & 2.25$\pm$0.09 & (4.45$\pm$0.17)e$-$15 &  6.3$\pm$0.3 & (2.88$\pm$0.17)e$-$15 &  9.7$\pm$0.6 \\
2004 May 23 & (4.50$\pm$0.11)e$-$14 & (1.76$\pm$0.11)e$-$14 & 2.56$\pm$0.17 & (4.02$\pm$0.76)e$-$15 & 11.2$\pm$2.1 & (2.60$\pm$0.99)e$-$15 & 17.3$\pm$6.6 \\
2006 Jan  9 & (3.86$\pm$0.22)e$-$14 & (1.79$\pm$0.11)e$-$14 & 2.16$\pm$0.18 & (5.22$\pm$0.73)e$-$15 &  7.4$\pm$1.1 & (4.44$\pm$0.93)e$-$15 &  8.7$\pm$1.9 \\
2006 Jun 12 & (6.49$\pm$0.15)e$-$14 & (2.65$\pm$0.19)e$-$14 & 2.45$\pm$0.19 & (4.89$\pm$0.55)e$-$15 & 13.3$\pm$1.5 & (3.79$\pm$0.55)e$-$15 & 17.1$\pm$2.5 \\
2007 Jan 24 & (5.36$\pm$0.10)e$-$14 & (1.96$\pm$0.07)e$-$14 & 2.74$\pm$0.11 & (4.23$\pm$0.53)e$-$15 & 12.7$\pm$1.6 & (2.73$\pm$0.29)e$-$15 & 19.6$\pm$2.1 \\
\hline
Observation & \SIII~$\lambda$9531 & \SiVI~1.965$\mu$m & \SIII/ & \SiX~1.430$\mu$m & \SIII/ & Pa$\beta$ & \SIII/ \\
Date & (erg/s/cm$^2$) & (erg/s/cm$^2$) & \SiVI & (erg/s/cm$^2$) & \SiX & (erg/s/cm$^2$) & Pa$\beta$ \\
\hline
2002 Apr 23 & (2.79$\pm$0.06)e$-$14 & (1.42$\pm$0.02)e$-$14 & 2.0$\pm$0.1 & (6.57$\pm$0.21)e$-$15 &  4.3$\pm$0.2 & (6.90$\pm$0.10)e$-$14 & 0.404$\pm$0.011 \\
2004 May 23 & (4.50$\pm$0.11)e$-$14 & (1.30$\pm$0.18)e$-$14 & 3.5$\pm$0.5 & (1.19$\pm$0.12)e$-$14 &  3.8$\pm$0.4 & (1.11$\pm$0.04)e$-$13 & 0.405$\pm$0.017 \\
2006 Jan  9 & (3.86$\pm$0.22)e$-$14 & (1.32$\pm$0.05)e$-$14 & 2.9$\pm$0.2 & (3.38$\pm$0.55)e$-$15 & 11.4$\pm$2.0 & (1.26$\pm$0.03)e$-$13 & 0.306$\pm$0.020 \\
2006 Jun 12 & (6.49$\pm$0.15)e$-$14 & (2.27$\pm$0.07)e$-$14 & 2.9$\pm$0.1 & (5.86$\pm$0.45)e$-$15 & 11.1$\pm$0.9 & (1.68$\pm$0.05)e$-$13 & 0.386$\pm$0.015 \\
2007 Jan 24 & (5.36$\pm$0.10)e$-$14 & (1.01$\pm$0.04)e$-$14 & 5.3$\pm$0.2 & (3.61$\pm$0.23)e$-$15 & 14.9$\pm$1.0 & (1.37$\pm$0.02)e$-$13 & 0.391$\pm$0.010 \\
\hline
\end{tabular}
\end{table*}

\begin{table*}
\caption{\label{optlines} 
Optical emission line fluxes and ratios}
\begin{tabular}{lccccccccc}
\hline
Observation & \OIII~$\lambda$5007 & \OIII~$\lambda$4959 & \OIII/ & \FeVII~$\lambda$3759 & \OIII/ & \FeVII~$\lambda$5159 & \OIII/ \\
Date & (erg/s/cm$^2$) & (erg/s/cm$^2$) & \OIII & (erg/s/cm$^2$) & \FeVII & (erg/s/cm$^2$) & \FeVII \\
\hline
2002 Mar  6 & (4.72$\pm$0.06)e$-$13 & (1.51$\pm$0.06)e$-$13 & 3.13$\pm$0.13 & (1.94$\pm$0.24)e$-$14 & 24.3$\pm$3.1 & (1.11$\pm$0.20)e$-$14 & 42.5$\pm$7.5 \\
2004 May 24 & (5.21$\pm$0.05)e$-$13 & (1.90$\pm$0.07)e$-$13 & 2.74$\pm$0.10 & (1.94$\pm$0.20)e$-$14 & 26.9$\pm$2.8 & (1.29$\pm$0.14)e$-$14 & 40.4$\pm$4.4 \\
2006 Jan  9 & (5.89$\pm$0.02)e$-$13 & (2.01$\pm$0.03)e$-$13 & 2.93$\pm$0.05 & (1.37$\pm$0.13)e$-$14 & 43.0$\pm$4.1 & (8.31$\pm$1.01)e$-$15 & 70.9$\pm$8.6 \\
2006 Jun 20 & (4.19$\pm$0.04)e$-$13 & (1.40$\pm$0.02)e$-$13 & 2.99$\pm$0.06 & (1.85$\pm$0.19)e$-$14 & 22.7$\pm$2.3 & (9.03$\pm$1.00)e$-$15 & 46.4$\pm$5.2 \\
2007 Feb 17 & (2.97$\pm$0.03)e$-$13 & (9.71$\pm$0.24)e$-$14 & 3.06$\pm$0.08 & (8.76$\pm$1.24)e$-$15 & 33.9$\pm$4.8 & (5.04$\pm$0.71)e$-$15 & 58.9$\pm$8.3 \\
\hline
Observation & \OIII~$\lambda$5007 &  \FeVII~$\lambda$5721 & \OIII/ & \FeVII~$\lambda$6087 & \OIII/ & H$\alpha$ & \OIII/ \\
Date & (erg/s/cm$^2$) & (erg/s/cm$^2$) & \FeVII & (erg/s/cm$^2$) & \FeVII & (erg/s/cm$^2$) & H$\alpha$ \\
\hline
2002 Mar  6 & (4.72$\pm$0.06)e$-$13 & (1.63$\pm$0.15)e$-$14 & 29.0$\pm$2.7 & (3.19$\pm$0.33)e$-$14 & 14.8$\pm$1.5 & (1.77$\pm$0.02)e$-$12 & 0.267$\pm$0.004 \\
2004 May 24 & (5.21$\pm$0.05)e$-$13 & (1.49$\pm$0.15)e$-$14 & 35.0$\pm$3.5 & (2.00$\pm$0.13)e$-$14 & 26.1$\pm$1.8 & (2.00$\pm$0.04)e$-$12 & 0.261$\pm$0.006 \\
2006 Jan  9 & (5.89$\pm$0.02)e$-$13 & (1.02$\pm$0.06)e$-$14 & 57.8$\pm$3.6 & (1.70$\pm$0.07)e$-$14 & 34.7$\pm$1.4 & (1.57$\pm$0.02)e$-$12 & 0.375$\pm$0.005 \\
2006 Jun 20 & (4.19$\pm$0.04)e$-$13 & (7.30$\pm$0.59)e$-$15 & 57.4$\pm$4.7 & (1.65$\pm$0.10)e$-$14 & 25.4$\pm$1.5 & (1.30$\pm$0.02)e$-$12 & 0.322$\pm$0.005 \\
2007 Feb 17 & (2.97$\pm$0.03)e$-$13 & (4.29$\pm$0.43)e$-$15 & 69.2$\pm$7.0 & (8.04$\pm$0.90)e$-$15 & 36.9$\pm$4.1 & (1.04$\pm$0.01)e$-$12 & 0.286$\pm$0.005 \\
\hline
\end{tabular}
\end{table*}

\begin{table*}
\caption{\label{ircont} 
Near-IR continuum fluxes and optical coronal line ratios}
\begin{tabular}{lccccccc}
\hline
Observation & $\lambda f_{1\mu{\rm m}}$ & $\lambda f_{1\mu{\rm m}}$/ & $\lambda f_{2.1\mu{\rm m}}$ & $\lambda f_{2.1\mu{\rm m}}$/ & $T_{\rm dust}$ & \FeVII~$\lambda$6087/ & \FeVII~$\lambda$5159/ \\
Date & (erg/s/cm$^2$) & \SIII & (erg/s/cm$^2$) & \SIII & (K) & \FeVII~$\lambda$3759 & \FeVII~$\lambda$6087 \\
\hline
2002 Apr 23 & (7.74$\pm$0.08)e$-$12 & 277.4$\pm$6.7  & (1.38$\pm$0.01)e$-$11 & 494.6$\pm$10.9 & 1371 & 1.644$\pm$0.268 & 0.348$\pm$0.071 \\
2004 May 23 & (1.48$\pm$0.07)e$-$11 & 328.9$\pm$17.1 & (1.85$\pm$0.02)e$-$11 & 411.1$\pm$11.5 & 1572 & 1.031$\pm$0.129 & 0.645$\pm$0.082 \\
2006 Jan  9 & (2.38$\pm$0.09)e$-$11 & 616.6$\pm$43.2 & (2.68$\pm$0.04)e$-$11 & 694.3$\pm$41.7 & 1730 & 1.241$\pm$0.128 & 0.489$\pm$0.062 \\
2006 Jun 12 & (2.32$\pm$0.06)e$-$11 & 357.5$\pm$11.8 & (3.56$\pm$0.02)e$-$11 & 548.5$\pm$13.2 & 1547 & 0.892$\pm$0.105 & 0.547$\pm$0.069 \\
2007 Jan 24 & (2.10$\pm$0.04)e$-$11 & 391.8$\pm$11.0 & (2.19$\pm$0.02)e$-$11 & 408.6$\pm$8.2  & 1291 & 0.918$\pm$0.166 & 0.627$\pm$0.112 \\
\hline
\end{tabular}
\end{table*}

We have five epochs of quasi-simultaneous (within less than two
months) near-IR and optical spectroscopy for NGC~5548 (see Tables
\ref{irlines} and \ref{optlines}). All data are of relatively high
signal-to-noise ratio (continuum $S/N \ga 50-100$). The near-IR
spectroscopy was obtained with the SpeX spectrograph \citep{Ray03} at
the NASA Infrared Telescope Facility (IRTF), a 3 m telescope on Mauna
Kea, Hawai'i, in the short cross-dispersed mode (SXD, $0.8-2.4$
$\mu$m). All data were obtained through a slit of $0.8\times15''$
giving an average spectral resolution of full width at half maximum
(FWHM) $\sim 400$~km~s$^{-1}$. The four epochs spanning the years
$2004 - 2007$ are our own data and were discussed and presented in
\citet{L08a} and \citet{L11a}. The near-IR spectrum from 2002 was
discussed by \citet{Rif06}. The optical spectra were obtained with the
FAST spectrograph \citep{Fast98} at the Tillinghast \mbox{1.5 m}
telescope on Mt. Hopkins, Arizona, using the 300~l/mm grating and a
$3''$ long-slit. This set-up resulted in a wavelength coverage of
$\sim 3720-7515$~\AA~and an average spectral resolution of FWHM $\sim
330$~km~s$^{-1}$. The slit was rotated to the parallactic angle for
all observations but the May 2004 epoch. However, the latter spectrum
was observed at a very low air mass ($\sec~z \sim 1.01$). The optical
data were discussed and presented in \citet{L08a} and \citet{L11a},
with the exception of the May 2004 spectrum, which was retrieved from
the FAST archive.

We have measured the fluxes of the strongest near-IR and optical
coronal lines and estimated their $1\sigma$ uncertainties as detailed
in \citet{L15}. In the near-IR, we have measured two sulfur lines and
two silicon lines, namely, \SVIII~$\lambda$9911, \SIX~1.252$\mu$m,
\SiVI~1.965$\mu$m and \SiX~1.430$\mu$m (see Table \ref{irlines}). In
the optical, we have measured four iron emission lines, namely,
\FeVII~$\lambda$3759, \FeVII~$\lambda$5159, \FeVII~$\lambda$5721 and
\FeVII~$\lambda$6087 (see Table \ref{optlines}). The $1\sigma$
uncertainties are $\sim 1-5\%$ for the strongest lines and $\sim
2-20\%$ for the weakest ones.
Since the spectra were obtained in non-photometric sky conditions, we
study in the following the temporal changes of the coronal lines in
relative rather than absolute flux. In particular, we scale the
coronal line emission to that of a strong, forbidden {\it
  low-ionisation} emission line that is unblended and observed in the
same spectrum. The emission region that produces the low-ionisation
narrow lines is believed to be located at large enough distances from
the central ionising source \citep[e.g.][]{Pogge88, Tad89} and to have
a relatively low density for its flux to remain constant on timescales
of decades. We have scaled the near-IR and optical coronal lines to
the \SIII~$\lambda$9531 and \OIII~$\lambda$5007 emission lines,
respectively.

In order to further constrain the significance of the observed coronal
line variability, we have considered the two extreme cases of where no
variability and the highest variability are expected. For both scaling
lines we observe also the other emission line that is emitted from the
same upper level, namely, \SIII~$\lambda 9069$ and \OIII~$\lambda
4959$.  Their observed ratios, which should be close to the
theoretical values of \SIII~$\lambda 9531$/$\lambda 9069$=2.58 and
\OIII~$\lambda 5007$/$\lambda 4959$=2.92 \citep{NIST}, are not
expected to vary and so their observed variability sets a lower
threshold for the significance of the coronal line variability. Then,
we have measured the fluxes of the two prominent broad emission lines
Pa$\beta$ (in the near-IR) and H$\alpha$ (in the optical). Since the
BELR is expected to be the most variable of any AGN emission line
region, the observed flux {\it increase} of these broad lines gives an
estimate of the maximum value that can be reached within the current
data set. These results are also listed in Tables \ref{irlines} and
\ref{optlines}.

In addition to the emission lines, we have measured in the near-IR
spectra the continuum fluxes at the rest-frame wavelengths of $\sim
1~\mu$m and $\sim 2.1~\mu$m (see Table \ref{ircont}). As we have shown
in \citet{L11a}, the former is dominated by the accretion disc flux,
which is believed to be the main source of ionising radiation in AGN
and so the driver of the observed variability, whereas the latter is
emitted from the hot dust component of the obscuring
torus. Furthermore, we have derived the hot dust temperature from
blackbody fits to the near-IR spectral continuum as described in
\citet{L11a} and list it also in Table \ref{ircont}. In Section
\ref{location}, we will compare these values to the optical coronal
line ratios \FeVII~$\lambda$6087/$\lambda$3759 and
\FeVII~$\lambda$5159/$\lambda$6087, which are suitable indicators of
the gas temperature and density \citep{Nuss82, Keen87}, respectively
(listed in Table \ref{ircont}).

\subsection{X-ray spectroscopy} \label{xray}

\begin{table*}
\caption{\label{xraylines} 
X-ray emission line and continuum fluxes}
\begin{tabular}{lccc}
\hline
Observation & \multicolumn{2}{c}{\OVII~f~0.561~keV$^\star$} & $f_{\rm 0.2-10keV}$ \\
Date & Ebrero et al. & Detmers et al. & Ebrero et al. \\ 
& (ph/s/cm$^2$) & (ph/s/cm$^2$) & (erg/s/cm$^2$) \\
\hline
2002 Jan 18 & (0.83$\pm$0.07)e$-$04 & (0.75$\pm$0.07)e$-$04 & (1.51$\pm$0.02)e$-$10 \\
2005 Apr 15 & (0.42$\pm$0.07)e$-$04 & (0.35$\pm$0.06)e$-$04 & (1.90$\pm$0.06)e$-$11 \\
2007 Aug 14 & (0.34$\pm$0.07)e$-$04 & (0.27$\pm$0.06)e$-$04 & (1.05$\pm$0.06)e$-$11 \\
\hline
\end{tabular}

\parbox[]{9.8cm}{$^\star$ Ebrero et al. contrary to Detmers et
  al. assume that the narrow emission lines are absorbed by the warm
  absorber.}

\end{table*}

As one of the brightest Seyfert 1 galaxies at X-ray frequencies,
NGC~5548 was intensively studied in the years 1999-2002, after which
period the source became unobservable by {\it XMM-Newton} and had only
limited observability with {\it Chandra}. Due to the limited
observability, there are two shorter {\it Chandra} observations in
2005 and 2007, until a long observational campaign took place in the
summer of 2013 and winter of 2013/14. Between 2002 and 2007, the time
period overlapping with our near-IR and optical spectroscopy, there
are a total of three {\it Chandra} observations of NGC~5548 with the
low-energy transmission grating in combination with the
high-resolution camera spectrograph (LETGS). The 2002 observations
showed the source in an average flux state, while the 2005 and
especially the 2007 spectra were obtained in low-flux states. The
exposure times were 340~ks, 141~ks and 162~ks in 2002, 2005, and 2007,
respectively. All three X-ray observations were taken over a period of
three days during two consecutive orbits. The low flux state helps in
the detection of emission lines but the shorter exposure times lead to
low S/N spectra.

The X-ray spectra obtained between 2002 and 2007 were analysed by
\cite{KS05}, \citet{Det08, Det09} and, most recently, Ebrero et
al. (2015, in prep.). Both \cite{Det09} and Ebrero et al. analyse all
previously observed X-ray spectra using the latest atomic data. The
difference between the two is that Ebrero et al. focus on detecting
changes in the warm absorber with continuum flux level and so,
contrary to \cite{Det09}, assume that the narrow emission lines are
absorbed by the warm absorber. Generally, it has been assumed that the
warm absorber does not affect the flux of the narrow emission lines,
however, Whewell et al. (2015, submitted) have recently presented
evidence that at least one and potentially up to three warm absorber
components do indeed absorb at least the \OVII~triplet and most likely
all narrow X-ray emission lines. In Table \ref{xraylines}, we list the
measurements and their 1$\sigma$ errors from both studies, Ebrero et
al. and Detmers et al., which yield similar results. Due to the still
rather high continuum level in 2005 and 2007 and lower S/N in 2005,
only the \OVII~f narrow emission line was detected in all three
spectra.

\section{The variability behaviour}

\subsection{The near-IR and optical coronal lines} \label{iroptvar}

\begin{figure}
\centerline{
\includegraphics[clip=true, bb=15 295 590 720, scale=0.45]{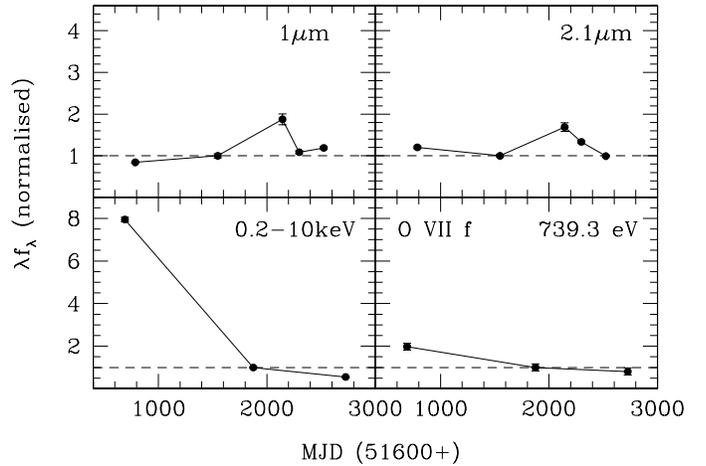}
}
\caption{\label{contvar} The variability of the continuum fluxes at
  rest-frame wavelengths of $\sim 1~\mu$m (sampling the accretion
  disc), $\sim 2.1~\mu$m (sampling the hot dust emission) and in the
  energy range $0.2-10$~keV (related to the accretion disc) and that
  of the \OVII~f X-ray coronal line. The near-IR continuum fluxes were
  divided by the \SIII~$\lambda 9531$ line flux and normalised to the
  value of the May 2004 epoch. The X-ray measurements were normalised
  to the value of the April 2005 epoch. Note the different y-axis
  scale in the bottom two panels. We plot $1\sigma$ error bars.}
\end{figure}

\begin{figure}
\centerline{
\includegraphics[scale=0.45]{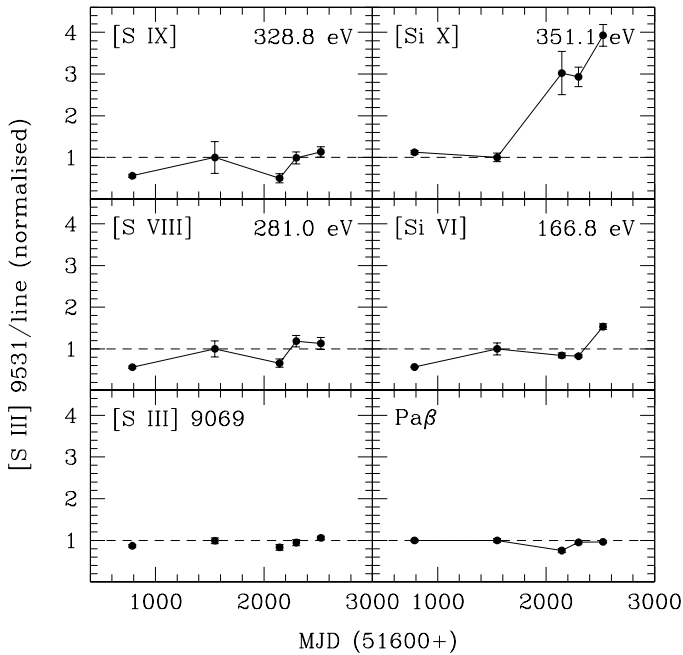}
}
\caption{\label{irvar} The variability of the near-IR coronal lines in
  the period April 2002 - January 2007. For comparison, we show also
  the \SIII~$\lambda 9069$ narrow line and Pa$\beta$ broad line, which
  are expected not to vary and to vary maximally, respectively. The
  \SIII~$\lambda 9531$/\SIII~$\lambda 9069$ flux ratios were
  normalised to the theoretical value, whereas those between
  \SIII~$\lambda 9531$ and the other lines were normalised to the
  value of the May 2004 epoch. We plot $1\sigma$ error bars. The
  ionisation potentials of the coronal lines are given at the top
  right.}
\end{figure}

\begin{figure}
\centerline{ 
\includegraphics[scale=0.45]{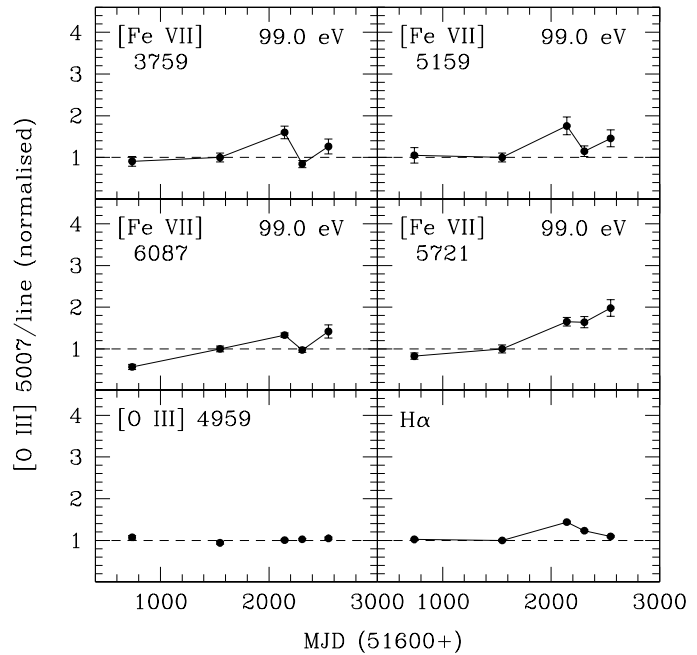}
}
\caption{\label{optvar} Same as Fig. \ref{irvar} for the optical
  coronal lines. For comparison, we show also the \OIII~$\lambda 4959$
  narrow line and H$\alpha$ broad line, which are expected not to vary
  and to vary maximally, respectively.}
\end{figure}

Over the $\sim 5$~year period sampled by the near-IR and optical
spectroscopy the ionising flux of the accretion disc, which is assumed
to be the main driver for the variability of the broad line, coronal
line and hot dust emission regions, stayed constant in the first two
years, increased by a factor of $\sim 2$ in the following two years
and decreased again to its original flux state during the last year
(see Fig. \ref{contvar}). We note that the flux increase of $\sim
19\%$ observed between the April 2002 and May 2004 epochs is
consistent with zero, once the deviations from the theoretical value
of the \SIII~$\lambda 9531$/\SIII~$\lambda 9069$ ratios measured in
these spectra are taken into account ($\sim 14.8\%$ and $\sim 1.1\%$,
respectively). The variability of the ionising continuum is mirrored
in that of the hot dust emission and Pa$\beta$ broad line region (see
Figs. \ref{contvar} and \ref{irvar}, respectively); also their flux
stayed constant in the first two years, increased in the following two
years (by $\sim 70\%$ and $\sim 30\%$, respectively), and then
continuously decreased to its original level for the next year. The
flux of the H$\alpha$ broad line region also stayed constant in the
first two years, but then decreased in the next two years (by $\sim
40\%$), followed by an increase to its original level during the last
year.

The variability response of the near-IR coronal lines was stronger
than that of the hot dust emission and broad emission line region and
in general very different from it. Whereas the flux of the \SIX~and
\SiX~lines, the lines with the highest ionisation potentials, stayed
constant in the first two years, the flux of the other two near-IR
coronal lines decreased by a factor of $\sim 2$. After this initial
decrease, the flux of the \SVIII~line stayed constant, whereas the
flux of the \SiVI~line stayed constant in the following two years, but
decreased further in the last year, resulting in a total decrease
during the observing period of a factor of $\sim 3$. The flux of the
\SIX~line decreased significantly (by a factor of $\sim 2$) only
during the last year of the observing period, whereas the \SiX~line
started to diminish already in May 2004 and by the end of the
observing period reached a flux decrease of a factor of $\sim 4$.

Similarly to the \SVIII~and \SiVI~near-IR coronal lines, the flux of
the optical coronal line \FeVII~$\lambda 6087$ decreased by $\sim
80\%$ in the first two years. whereas the flux of the other three
\FeVII~lines stayed constant. The flux of the \FeVII~$\lambda 6087$
line further decreased in the following two years (by $\sim 30\%$),
then increased and decreased again, thus reaching a flux decrease of a
factor of $\sim 2$ by the end of the observing period. Similarly to
the H$\alpha$ broad line, the flux of the \FeVII~$\lambda 5159$~line,
after staying constant for the first two years, decreased in the
following two years (by $\sim 80\%$), followed by an increase to its
original level during the last year. Similarly to the \SiX~near-IR
coronal line, the flux of the \FeVII~$\lambda 5721$~line started to
diminish in May 2004 and by the end of the observing period reached a
flux decrease of a factor of $\sim 2$. Finally, after May 2004, the
\FeVII~$\lambda 3759$~line varied similarly to the \FeVII~$\lambda
6087$~line; a flux decrease by $\sim 60\%$ followed by a flux increase
(by $\sim 90\%$) and then flux decrease (by $\sim 50\%$).

\subsection{The X-ray coronal lines} \label{xrayvar}

In the $\sim 5$~year period sampled by the X-ray spectroscopy the
unabsorbed $0.2-10$~keV continuum flux, which is assumed to be
produced by the central ionising source and so to be linked to the
accretion disc flux, decreased by a factor of $\sim 8$ in the first
three years, and then decreased further by a factor of $\sim 2$ in the
next two years (see Fig. \ref{contvar}).

The variability response of the forbidden X-ray coronal line \OVII~f
was similar to that of the near-IR and optical coronal lines (see
Fig. \ref{contvar}). Its line flux decreased in the first three years
by a factor of $\sim 2$ and stayed constant in the next two years. We
note that this result is independent of the assumption of whether the
X-ray narrow lines are affected by the warm absorber, since the ratio
between the absorbed and unabsorbed line flux is similar for all three
observing epochs. \cite{Det08} list also the flux or non-detection
limit for the \NeIX~f line, which is detected only in the 2002
spectrum. The trend for this much weaker X-ray emission line is
similar to that observed for the \OVII~f line; a decrease in flux
during the observing period by a factor of $\ga 3$. Finally, it is
worth noting that the flux of both the \OVII~f line and the unabsorbed
X-ray continuum recovered by the summer of 2013 to a level only
slightly below that observed in 2002 \citep{Kaastra14}.

\section{The origin of the coronal line emission region} \label{location}

The current understanding is that the coronal line region in AGN is
dust-free, photoionised and located beyond the BELR at distances from
the central ionising source similar to those of the hot inner face of
the obscuring dusty torus. In \citet{L15} we confirmed for the source
NGC~4151 that the coronal line region is photoionised but showed that
it is located well beyond the hot dusty torus, at a distance from the
central ionising source of a few light-years. This explained the low
variability amplitude observed for the coronal lines in this source.

The situation in the source NGC~5548 seems at first glance to be very
different. Whereas the broad emission lines changed only little (by
$\sim 30-40\%$) in the period covered by our data and the hot dust
emission changed moderately (by $\sim 70\%$), the coronal lines varied
strongly (by factors of $\sim 2-4$). However, the observed high
variability amplitude of the coronal lines is mainly due to a flux
decrease. In fact, none of the coronal lines reacted with a flux
increase to the increase in ionising flux observed between the May
2004 and January 2006 epochs as the Pa$\beta$ broad line and hot dust
emission did. 

In the following, we show that the coronal line region in NGC~5548 is
similar to that in NGC~4151, i.e. it is a separate entity, most likely
an X-ray heated wind, that has a low density and a high ionisation
parameter and that is located far away from the dusty torus (Section
\ref{plasma}). However, this wind appears to be stronger in NGC~5548
than in NGC~4151 and so, most likely, it undergoes a more severe
adiabatic expansion and cooling after being launched (Section
\ref{adiabatic}). Finally, in Section \ref{tidal}, we discuss the
source NGC~5548 in the context of stellar tidal disruption events,
which have been proposed as an alternative explanation for the fact
that strong coronal line emitters often show a strong fading of their
coronal lines over a period of several years.

\subsection{Plasma diagnostics and the similarity between NGC 5548 and NGC 4151} \label{plasma}

\begin{figure}
\centerline{
\includegraphics[scale=0.43]{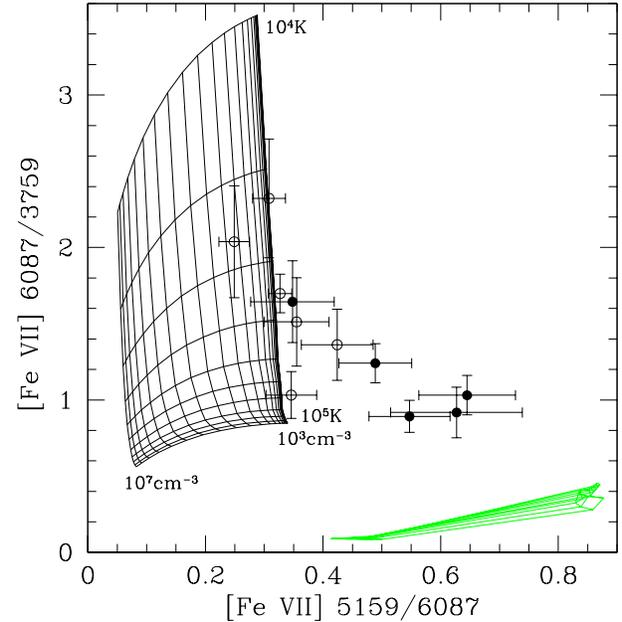}
}
\caption{\label{FeVIItemp} The observed optical coronal line ratios
  \FeVII~$\lambda$6087/$\lambda$3759 versus
  \FeVII~$\lambda$5159/$\lambda$6087 for NGC~5548 (filled circles) and
  NGC~4151 (open circles), overlaid with curves of constant
  temperature ($\log T=4, 4.1, 4.2, ... 5$~K) and constant number
  density ($\log n_e=3, 3.2, 3.4, ... 7$~cm$^{-3}$) for the case of
  photoionisation equilibrium. The case of collisional ionisation
  equilibrium is shown in green. We plot $1\sigma$ error bars.}
\end{figure}

\begin{figure}
\centerline{
\includegraphics[scale=0.43]{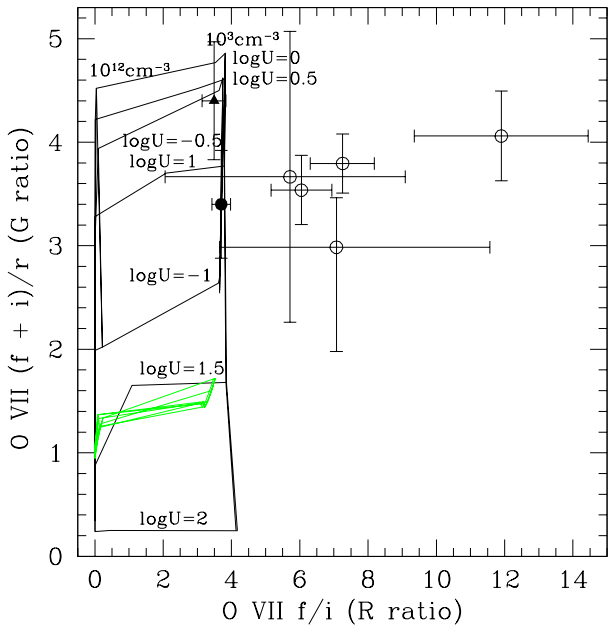}
}
\caption{\label{OVIItemp} The observed X-ray coronal line ratios
  \OVII~(f$+$i)/r (G ratio) versus \OVII~f/i (R ratio) for NGC~5548
  (filled symbols) and NGC~4151 (open circles), overlaid with curves
  of constant ionisation parameter ($\log U=-1, -0.5, 0, ... 2$) for
  two constant number densities ($n_e=10^3$~cm$^{-3}$ and
  $n_e=10^{12}$~cm$^{-3}$) and a fixed column density of $N_{\rm
    H}=10^{21}$~cm$^{-2}$, assuming photoionisation equilibrium. For
  NGC~5548 we plot both the absorbed (filled circle) and unabsorbed
  case (filled triangle). The case of collisional ionisation
  equilibrium is shown in green. We plot $1\sigma$ error bars.}
\end{figure}

As in \citet{L15}, we estimate the density of the coronal line gas
using the three optical iron lines \FeVII~$\lambda$3759,
\FeVII~$\lambda$5159 and \FeVII~$\lambda$6087, since, as discussed by
previous studies, the line ratios \FeVII~$\lambda$6087/$\lambda$3759
and \FeVII~$\lambda$5159/$\lambda$6087 are suitable indicators of
temperature and density, respectively \citep{Nuss82, Keen87}. For
these emission lines, we have generated a temperature versus density
grid using version 13.03 of the plasma simulation code
\cloudy~\citep[last described by][]{newCloudy} (see
Fig. \ref{FeVIItemp}). For the photoionisation simulations we have
assumed solar abundances as given by \citet{Grev10} and approximated
the incident radiation field with the mean AGN spectral energy
distribution (SED) derived by \citet{Math87}. Recently, \citet{Meh15a}
observed the hard X-ray portion of the SED and found that it continues
to rise, in $\nu f_\nu$, until at least $\sim 80$~keV. So NGC~5548 was
in a very hard state when their observations were done. Very hard
X-rays actually have only modest effects on the physical conditions
within a photoionised cloud due to their small photoelectric cross
section. The emission line spectrum is mainly affected by the FUV/XUV
part of the SED, which is not directly observable except in the very
highest redshift objects \citep{Colli15}. The FUV/XUV part of the
older SED was derived from line observations which were sensitive to
that portion of the spectrum. The two SEDs are in qualitative
agreement in this part of the spectrum, so our results would not have
changed had we used the newest observations. We have assumed the cases
of either photoionisation or collisional ionisation equilibrium.

The measurements for NGC~5548 (filled circles) give a similar answer
to those for NGC~4151 (open circles); the coronal line gas is
photoionised rather than collisionally ionised and its density appears
to be relatively low. All but one observing epochs for NGC~5548
constrain it to $n_e \sim 10^3$~cm$^{-3}$ within $\sim 3\sigma$,
whereas the May 2004 epoch reaches this value within $\sim
4\sigma$. We note that the estimated density is several orders of
magnitude below the critical density of \FeVII~of $n_{\rm crit} \sim 3
\times 10^7$~cm$^{-3}$. We find again for NGC~5548, as we did
previously for NGC~4151, that all measured \FeVII~ratios lie to the
right of the theoretical grids. This indicates most likely a large
ionisation parameter, since in such a case the contribution from
fluorescence increases significantly the
\FeVII~$\lambda$5159/$\lambda$6087 ratio \citep[see Fig. 7 in][]{L15}.

As in \citet{L15}, we estimate the ionisation parameter of the coronal
line gas using the three X-ray lines from the helium-like ion of
oxygen \OVII~f, \OVII~i and \OVII~r. Using the plasma simulation code
\cloudy, we have generated a ionisation parameter versus density grid
for the \OVII~lines (see Fig. \ref{OVIItemp}). As discussed by
previous studies, the X-ray line ratios \OVII~(f$+$i)/r (the so-called
G ratio) and \OVII~f/i (the so-called R ratio) trace the ionisation
parameter, which is directly related to the kinetic gas temperature,
and density, respectively, for a given column density \citep{Por00,
  Porter07}. We assumed again the cases of either photoionisation or
collisional ionisation equilibrium. Neither of the studies mentioned
in Section \ref{xray} detects all three \OVII~lines and, therefore, no
density and/or ionisation parameter can be determined from this
triplet for the period 2002-2007. However, the entire \OVII~triplet is
detected by \cite{Kaastra14} when the source was in an obscured state
similar to what is generally observed for NGC~4151. \cite{Kaastra14}
assumed that the narrow emission lines are not absorbed by the warm
absorber \citep[similar to the study of][]{Det09} and from their data
we determine a G ratio of 4.40$\pm$0.57 and an R ratio of
3.49$\pm$0.35 (filled triangle in Fig. \ref{OVIItemp}). \cite{And10}
combined all {\it Chandra} spectra taken before 2007 and measure
the~\OVII~triplet in the stacked HETG and LETG spectra, also assuming
that the lines are not absorbed by the warm absorber. From their data
we calculate a G ratio of 8.8$^{+6.2}_{-6.1}$ and 4.6$^{+2.8}_{-1.4}$
and an R ratio of 3.3$^{+1.5}_{-1.4}$ and 2.8$^{+1.5}_{-0.6}$ for the
HETG and LETG data, respectively. Given the large errors, these
results are consistent with the {\it XMM-Newton} results of
\cite{Kaastra14}. Therefore, we will use here the latter data given
the higher quality, although the observations are not contemporaneous
with our optical spectroscopy. If we assume that the \OVII~triplet is
affected by the warm absorber, both the G and R ratios decrease,
because the resonance and intercombination lines are absorbed whereas
the forbidden line is not. \citet{Meh15b} have recently pointed out
that inner shell lines of \OVI~can absorb components of the
\OVII~lines used as a diagnostic here and in observations of NGC~4151
presented in \citet{L15}. This hypothesis is attractive because it is
hard to think of any other way to account for the unusual ratios found
in our previous paper. But the problem with their explanation in the
case of NGC~5548 is that the measured \OVI~column density
\citep[$N_{\rm OVI}=10^{16}$~cm$^{-2}$;][]{KS05} of the warm absorber,
which potentially can be cospatial with the \OVII~line emitting
region, is $\sim1.5$~dex smaller than that needed to produce the
absorption assuming their atomic data. Then, using the line strengths
from Whewell et al. (2015, submitted), the G and R ratios become
3.40$\pm$0.52 and 3.70$\pm$0.27, respectively (filled circle in
Fig. \ref{OVIItemp}). The absorbed values constrain the ionisation
parameter of the coronal line gas in NGC~5548 to $\log U \sim 1$, as
found previously for NGC~4151.

Then, using the unabsorbed X-ray luminosity of $L_{\rm 2-10keV} \sim 4
\times 10^{43}$~erg~s$^{-1}$ of the highest-flux epoch as a proxy for
the ionising luminosity producing \OVII~and the best-fit X-ray
spectral slope of this epoch of $\Gamma \sim 1.88$ \citep{KS05} to
estimate the mean ionising photon energy, the calculated distance of
the coronal line region in NGC~5548 from the central ionising source
for the above gas density is $\tau_{\rm lt} \sim 2850$~light
days~$\sim 7.8$~light years. As found previously for NGC~4151, this
value places this region well beyond the hot inner face of the
obscuring dusty torus, which in NGC~5548 is measured by dust
reverberation campaigns to be $\sim 40-60$~days
\citep{Kosh14}. Furthermore, it is consistent with the fact that a
recovery of the flux of the coronal lines to the high level seen in
the 2002 epoch is occurring only in 2013 (see Section \ref{xrayvar})
and that only a flux decrease is observed in the 5-yr time period
spanned by our own data set.

\subsection{The X-ray heated wind scenario and the difference between NGC~5548 and NGC~4151} \label{adiabatic}

\begin{figure}
\centerline{
\includegraphics[scale=0.42]{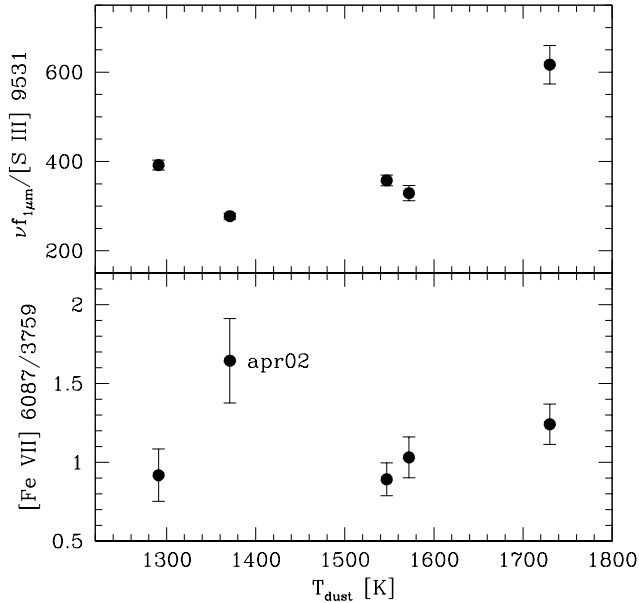}
}
\caption{\label{dusttemp} The top and bottom panels show the continuum
  flux at rest-frame wavelength of $\sim 1~\mu$m (sampling the
  accretion disc) scaled to the \SIII~$\lambda 9531$ line flux and the
  optical coronal line ratio \FeVII~$\lambda$6087/$\lambda$3759, which
  is a suitable indicator of the gas temperature (the higher its
  value, the lower the temperature; see Fig. \ref{FeVIItemp}),
  respectively, versus the hot dust blackbody temperature.}
\end{figure}

\citet{Pier95} proposed that the coronal line region is a layer on the
inner part of the dusty torus that becomes an efficient coronal line
emitter only when evaporated in an X-ray heated wind. In this
scenario, the wind will have undergone adiabatic expansion from its
launch location at the inner face of the torus until the gas density
and distance from the ionising source are optimal to give the required
high ionisation parameter. In the process, the coronal line gas will
have cooled. In Fig. \ref{dusttemp} we present tentative evidence that
this scenario applies to NGC~5548, similar to what we found previously
for NGC~4151. In the bottom panel, we compare the temperature of the
coronal line gas as measured by the line ratio
\FeVII~$\lambda$6087/\FeVII~$\lambda$3759 (the higher its value, the
lower the gas temperature; see Fig. \ref{FeVIItemp}) with that of the
hot dust. We find that the two temperatures behave in opposite ways;
the temperature of the coronal line gas is high when the hot dust
temperature is low and vice versa. Only the data from April 2002 is an
exception to this trend and might indicate a reddening event. In the
top panel of Fig. \ref{dusttemp}, we show the continuum flux at
rest-frame wavelength of $\sim 1$~$\mu$m, which samples the accretion
disc luminosity, versus the hot dust temperature. A clear correlation
is apparent, which indicates that the change in temperature for the
hot dust is due to direct heating by the central ionising
source. Therefore, the increased AGN radiation that heats the dusty
torus also increases the cooling of the coronal line gas.

In the scenario of \citet{Pier95}, the dusty clouds will be evaporated
in an X-ray heated wind more efficiently for higher AGN luminosities,
which will lead to both an increase in mass outflow rate and a
stronger adiabatic expansion. We observe this effect individually for
NGC~5548 and NGC~4151, but it seems to also apply when comparing the
two sources with each other. The coronal lines in NGC~5548 are a
factor of $\sim 3-5$ stronger than in NGC~4151 relative to the
low-ionisation narrow emission lines, and so are both the $\sim
1$~$\mu$m and $\sim 2.1$~$\mu$m luminosities (higher by a factor of
$\sim 3$ and $\sim 2$, respectively). This indicates that the X-ray
heated wind producing the coronal lines in NGC~5548 is stronger than
that in NGC~4151, in which case we also expect a stronger adiabatic
expansion. The much stronger fading of the coronal lines with time
observed for NGC 5548 relative to NGC~4151 is most likely a
manifestation of it.

\subsection{Strong coronal line emitters as stellar tidal disruption events} \label{tidal}

Such a strong variability of the coronal lines, i.e. fading of the
flux, as we observe for NGC~5548 has been reported so far for only one
AGN \citep[IC~3599;][]{Grupe95, Brandt95} and is usually associated
with a new class of non-active galaxies, the so-called strong coronal
line emitters. Several of these sources have recently been detected in
the SDSS \citep{Kom08, Gel09, Kom09, WangT11, WangT12, Yang13,
  Rose15}. A stellar tidal disruption event seems to be the most
plausible explanation for the strong fading of the coronal lines
observed in these sources over a time period of several years. The
tidal disruption of a star by a supermassive black hole produces an
accretion event, a flare, that subsequently fades on timescales of
several months to a year. The accretion flare is seen as an X-ray/UV
outburst, which can produce highly ionised emission lines, including
emission from \HeII. The X-ray luminosity outburst is expected to fade
with time as $t^{-5/3}$ or slightly shallower than this depending on
the type of the star \citep{Rees88, Phi89, Loda09} and subsequently a
strong fading of the high-ionisation lines is expected.

Could the strong coronal line variability that we observe for NGC~5548
have been caused by a stellar tidal disruption event? Such an event
was raised as a possibility for the strong increase (by a factor of
$\ga 10$) of the \HeII~$\lambda 4686$ emission line flux observed for
this source between January and May 1984 by \citet{Pet86} and
interpreted as the addition of a new gas component unrelated to the
BELR. The recently discovered strong coronal line emitters have line
ratios \FeVII~$\lambda 6087$/\OIII~$\lambda 5007$~$ \sim 0.1-1$ and
have \FeX~emission comparable to or stronger than \FeVII~before the
fading of the lines. In NGC~5548, we measure line ratios
\FeVII~$\lambda 6087$/\OIII~$\lambda 5007$~$ \sim 0.03-0.07$, which
puts this source in the vicinity of the strong coronal line emitters.
However, the \FeX~$\lambda 6375$ emission line is a factor of $\sim 5$
weaker than the \FeVII~$\lambda 6087$ emission line. As expected for
stellar tidal disruption events, the high-state of the coronal lines
in 2002 is accompanied by a high state of the X-ray flux, which
subsequently fades strongly. The observed X-ray flux decrease of a
factor of $\sim 8$ in 3 years roughly obeys the expected $t^{-5/3}$
decay law for these events (albeit based on only two points). However,
the flux of both the X-ray continuum and \OVII~f X-ray coronal line
are observed to have recovered by 2013 to their original high state
(see Section \ref{xrayvar}). Finally, the \HeII~emission line did not
show the continuous flux decline observed for most of the coronal
lines. Its flux varied significantly during the observing period, but
fluctuated between increase and decrease from epoch to epoch. Given
these considerations, it is therefore unlikely that the strong coronal
line variability that we observe for NGC~5548 was caused by a stellar
tidal disruption event.

\section{Summary and conclusions}

We have presented the second extensive study of the coronal line
variability in an AGN. Our data set for the well-studied source
NGC~5548 is unprecedented in that it includes five epochs of
quasi-simultaneous optical and near-IR spectroscopy spanning a period
of $\sim 5$~years and three epochs of X-ray spectroscopy overlapping
in time with it. Our main results are as follows.

\smallskip

(i) Whereas the broad emission lines changed only little (by $\sim
30-40\%$) and the hot dust emission changed moderately (by $\sim
70\%$), the coronal lines varied strongly (by factors of $\sim
2-4$). However, the observed high variability amplitude of the coronal
lines is mainly due to a flux decrease. In fact, none of the coronal
lines reacted with a flux increase to the increase in ionising flux
observed between the May 2004 and January 2006 epochs as the Pa$\beta$
broad line and hot dust emission did.

(ii) We have applied plasma diagnostics to the optical \FeVII~and
X-ray \OVII~emission lines in order to constrain the gas number
density, temperature and ionisation parameter of the coronal line
region. We find that this gas has a relatively low density of $n_e
\sim 10^3$~cm$^{-3}$ and requires a relatively high ionisation
parameter of $\log U \sim 1$, similar to our previous results for the
source NGC~4151. We estimate the distance of the coronal line region
in NGC~5548 from the central ionising source for the above gas density
to be $\tau_{\rm lt} \sim 7.8$~light years. This value puts this
region well beyond the hot inner face of the obscuring dusty torus
\citep[of $\sim 2$~light months;][]{Kosh14}.

(iii) The relatively high ionisation parameter at large distances make
it likely that the coronal line region is an independent entity. One
possibility is that it is an X-ray heated wind as first proposed by
\citet{Pier95}. As in the case of NGC~4151, we find support for this
scenario in the form of a temperature anti-correlation between the
coronal line gas and hot dust, which indicates that the increased AGN
radiation that heats the dusty torus appears to increase the cooling
efficiency of the coronal line gas, most likely due to a stronger
adiabatic expansion. The strong coronal line variability of NGC~5548
can also be explained within this picture; relative to the
low-ionisation narrow emission lines the coronal lines in NGC~5548 are
a factor of $\sim 3-5$ stronger than those in NGC~4151 indicating a
much stronger wind, in which case a stronger adiabatic expansion of
the gas and so fading of the line emission is expected.

(iv) Since NGC~5548 can be classified as a strong coronal line
emitter, we investigate if the strong coronal line variability could
have been caused by a stellar tidal disruption event, as is often
proposed for this kind of sources. Given that the flux of both the
X-ray continuum and \OVII~f X-ray coronal line are observed to have
recovered by 2013 to their original high state and that the
\HeII~emission line did not show the continuous flux decline observed
for most of the coronal lines, we consider this scenario an unlikely
explanation for the strong coronal line variability.

\section*{Acknowledgments}

HL is supported by a European Union COFUND/Durham Junior Research
Fellowship (under EU grant agreement number 267209). KCS thanks the
astronomy group at Durham University for its hospitality during a
collaborative visit. GJF acknowledges support by NSF (1108928,
1109061, and 1412155), NASA (10-ATP10-0053, 10-ADAP10-0073,
NNX12AH73G, and ATP13-0153), and STScI (HST-AR-13245, GO-12560,
HST-GO-12309, GO-13310.002-A, and HST-AR-13914), and to the Leverhulme
Trust for support via the award of a Visiting Professorship at Queen's
University Belfast (VP1-2012-025).

\bibliography{/Users/herminelandt/references}

\begin{thebibliography}{}

\bibitem[\protect\citeauthoryear{{Andrade-Vel{\'a}zquez}
  et~al.}{{Andrade-Vel{\'a}zquez} et~al.}{2010}]{And10}
{Andrade-Vel{\'a}zquez}, M., {Krongold}, Y., {Elvis}, M., {Nicastro}, F.,
  {Brickhouse}, N., {Binette}, L., {Mathur}, S.,  \& {Jim{\'e}nez-Bail{\'o}n},
  E. 2010, \apj, 711, 888

\bibitem[\protect\citeauthoryear{{Appenzeller} \& {Oestreicher}}{{Appenzeller}
  \& {Oestreicher}}{1988}]{App88}
{Appenzeller}, I.,  \& {Oestreicher}, R. 1988, \aj, 95, 45

\bibitem[\protect\citeauthoryear{{Brandt}, {Pounds}, \& {Fink}}{{Brandt}
  et~al.}{1995}]{Brandt95}
{Brandt}, W.~N., {Pounds}, K.~A.,  \& {Fink}, H. 1995, \mnras, 273, L47

\bibitem[\protect\citeauthoryear{{Collinson} et~al.}{{Collinson}
  et~al.}{2015}]{Colli15}
{Collinson}, J.~S., {Ward}, M.~J., {Done}, C., {Landt}, H., {Elvis}, M.,  \&
  {McDowell}, J.~C. 2015, \mnras, 449, 2174

\bibitem[\protect\citeauthoryear{{Detmers} et~al.}{{Detmers}
  et~al.}{2008}]{Det08}
{Detmers}, R.~G., {Kaastra}, J.~S., {Costantini}, E., {McHardy}, I.~M.,  \&
  {Verbunt}, F. 2008, \aap, 488, 67

\bibitem[\protect\citeauthoryear{{Detmers}, {Kaastra}, \& {McHardy}}{{Detmers}
  et~al.}{2009}]{Det09}
{Detmers}, R.~G., {Kaastra}, J.~S.,  \& {McHardy}, I.~M. 2009, \aap, 504, 409

\bibitem[\protect\citeauthoryear{{Erkens}, {Appenzeller}, \& {Wagner}}{{Erkens}
  et~al.}{1997}]{Erk97}
{Erkens}, U., {Appenzeller}, I.,  \& {Wagner}, S. 1997, \aap, 323, 707

\bibitem[\protect\citeauthoryear{Fabricant et~al.}{Fabricant
  et~al.}{1998}]{Fast98}
Fabricant, D., Cheimets, P., Caldwell, N.,  \& Geary, J. 1998, PASP, 110, 79

\bibitem[\protect\citeauthoryear{{Ferland} et~al.}{{Ferland}
  et~al.}{2013}]{newCloudy}
{Ferland}, G.~J., et~al. 2013, RMxAA, 49, 137

\bibitem[\protect\citeauthoryear{{Gelbord}, {Mullaney}, \& {Ward}}{{Gelbord}
  et~al.}{2009}]{Gel09}
{Gelbord}, J.~M., {Mullaney}, J.~R.,  \& {Ward}, M.~J. 2009, \mnras, 397, 172

\bibitem[\protect\citeauthoryear{{Grevesse} et~al.}{{Grevesse}
  et~al.}{2010}]{Grev10}
{Grevesse}, N., {Asplund}, M., {Sauval}, A.~J.,  \& {Scott}, P. 2010, Ap\&SS,
  328, 179

\bibitem[\protect\citeauthoryear{{Grupe} et~al.}{{Grupe}
  et~al.}{1995}]{Grupe95}
{Grupe}, D., {Beuermann}, K., {Mannheim}, K., {Bade}, N., {Thomas}, H.-C., {de
  Martino}, D.,  \& {Schwope}, A. 1995, \aap, 299, L5

\bibitem[\protect\citeauthoryear{{Kaastra} et~al.}{{Kaastra}
  et~al.}{2014}]{Kaastra14}
{Kaastra}, J.~S., et~al. 2014, Science, 345, 64

\bibitem[\protect\citeauthoryear{{Keenan} \& {Norrington}}{{Keenan} \&
  {Norrington}}{1987}]{Keen87}
{Keenan}, F.~P.,  \& {Norrington}, P.~H. 1987, \aap, 181, 370

\bibitem[\protect\citeauthoryear{{Kollatschny} et~al.}{{Kollatschny}
  et~al.}{2001}]{Kolla01}
{Kollatschny}, W., {Bischoff}, K., {Robinson}, E.~L., {Welsh}, W.~F.,  \&
  {Hill}, G.~J. 2001, \aap, 379, 125

\bibitem[\protect\citeauthoryear{{Komossa} et~al.}{{Komossa}
  et~al.}{2009}]{Kom09}
{Komossa}, S., et~al. 2009, \apj, 701, 105

\bibitem[\protect\citeauthoryear{{Komossa} et~al.}{{Komossa}
  et~al.}{2008}]{Kom08}
{Komossa}, S., et~al. 2008, \apjl, 678, L13

\bibitem[\protect\citeauthoryear{{Koshida} et~al.}{{Koshida}
  et~al.}{2014}]{Kosh14}
{Koshida}, S., et~al. 2014, \apj, 788, 159

\bibitem[\protect\citeauthoryear{Kramida et~al.}{Kramida et~al.}{2013}]{NIST}
Kramida, A., {Yu.~Ralchenko}, Reader, J.,  \& {and NIST ASD Team}. 2013, {NIST
  Atomic Spectra Database (ver. 5.1), [Online]. Available:
  {\tt{http://physics.nist.gov/asd}}. National Institute of Standards and
  Technology, Gaithersburg, MD.}

\bibitem[\protect\citeauthoryear{Landt et~al.}{Landt et~al.}{2008}]{L08a}
Landt, H., Bentz, M.~C., Ward, M.~J., Elvis, M., Peterson, B.~M., Korista,
  K.~T.,  \& Karovska, M. 2008, ApJS, 174, 282

\bibitem[\protect\citeauthoryear{Landt et~al.}{Landt et~al.}{2011}]{L11a}
Landt, H., Elvis, M., Ward, M.~J., Bentz, M.~C., Korista, K.~T.,  \& Karovska,
  M. 2011, MNRAS, 414, 218

\bibitem[\protect\citeauthoryear{Landt et~al.}{Landt et~al.}{2015}]{L15}
Landt, H., Ward, M.~J., Steenbrugge, K.~C.,  \& Ferland, G.~J. 2015, MNRAS,
  449, 3795

\bibitem[\protect\citeauthoryear{{Lodato}, {King}, \& {Pringle}}{{Lodato}
  et~al.}{2009}]{Loda09}
{Lodato}, G., {King}, A.~R.,  \& {Pringle}, J.~E. 2009, \mnras, 392, 332

\bibitem[\protect\citeauthoryear{{Mathews} \& {Ferland}}{{Mathews} \&
  {Ferland}}{1987}]{Math87}
{Mathews}, W.~G.,  \& {Ferland}, G.~J. 1987, \apj, 323, 456

\bibitem[\protect\citeauthoryear{{Mazzalay} et~al.}{{Mazzalay}
  et~al.}{2013}]{Maz13}
{Mazzalay}, X., {Rodr{\'{\i}}guez-Ardila}, A., {Komossa}, S.,  \& {McGregor},
  P.~J. 2013, \mnras, 430, 2411

\bibitem[\protect\citeauthoryear{{Mehdipour} et~al.}{{Mehdipour}
  et~al.}{2015}]{Meh15a}
{Mehdipour}, M., et~al. 2015, \aap, 575, A22

\bibitem[\protect\citeauthoryear{{Mehdipour}, {Kaastra}, \&
  {Raassen}}{{Mehdipour} et~al.}{2015}]{Meh15b}
{Mehdipour}, M., {Kaastra}, J.~S.,  \& {Raassen}, A.~J.~J. 2015, A\&A, 579, 87

\bibitem[\protect\citeauthoryear{{M{\"u}ller S{\'a}nchez} et~al.}{{M{\"u}ller
  S{\'a}nchez} et~al.}{2006}]{Mueller06}
{M{\"u}ller S{\'a}nchez}, F., {Davies}, R.~I., {Eisenhauer}, F., {Tacconi},
  L.~J., {Genzel}, R.,  \& {Sternberg}, A. 2006, \aap, 454, 481

\bibitem[\protect\citeauthoryear{{M{\"u}ller-S{\'a}nchez}
  et~al.}{{M{\"u}ller-S{\'a}nchez} et~al.}{2011}]{Mueller11}
{M{\"u}ller-S{\'a}nchez}, F., {Prieto}, M.~A., {Hicks}, E.~K.~S.,
  {Vives-Arias}, H., {Davies}, R.~I., {Malkan}, M., {Tacconi}, L.~J.,  \&
  {Genzel}, R. 2011, \apj, 739, 69

\bibitem[\protect\citeauthoryear{{Nussbaumer}, {Storey}, \&
  {Storey}}{{Nussbaumer} et~al.}{1982}]{Nuss82}
{Nussbaumer}, H., {Storey}, P.~J.,  \& {Storey}, P.~J. 1982, \aap, 113, 21

\bibitem[\protect\citeauthoryear{{Penston} et~al.}{{Penston}
  et~al.}{1984}]{Pen84}
{Penston}, M.~V., {Fosbury}, R.~A.~E., {Boksenberg}, A., {Ward}, M.~J.,  \&
  {Wilson}, A.~S. 1984, \mnras, 208, 347

\bibitem[\protect\citeauthoryear{{Peterson} \& {Ferland}}{{Peterson} \&
  {Ferland}}{1986}]{Pet86}
{Peterson}, B.~M.,  \& {Ferland}, G.~J. 1986, \nat, 324, 345

\bibitem[\protect\citeauthoryear{{Phinney}}{{Phinney}}{1989}]{Phi89}
{Phinney}, E.~S. 1989, in IAU Symposium, Vol. 136, The Center of the Galaxy,
  ed. M.~{Morris}, 543

\bibitem[\protect\citeauthoryear{{Pier} \& {Voit}}{{Pier} \&
  {Voit}}{1995}]{Pier95}
{Pier}, E.~A.,  \& {Voit}, G.~M. 1995, \apj, 450, 628

\bibitem[\protect\citeauthoryear{{Pogge}}{{Pogge}}{1988}]{Pogge88}
{Pogge}, R.~W. 1988, \apj, 328, 519

\bibitem[\protect\citeauthoryear{{Porquet} \& {Dubau}}{{Porquet} \&
  {Dubau}}{2000}]{Por00}
{Porquet}, D.,  \& {Dubau}, J. 2000, \aaps, 143, 495

\bibitem[\protect\citeauthoryear{{Porter} \& {Ferland}}{{Porter} \&
  {Ferland}}{2007}]{Porter07}
{Porter}, R.~L.,  \& {Ferland}, G.~J. 2007, \apj, 664, 586

\bibitem[\protect\citeauthoryear{{Prieto}, {Marco}, \& {Gallimore}}{{Prieto}
  et~al.}{2005}]{Prieto05}
{Prieto}, M.~A., {Marco}, O.,  \& {Gallimore}, J. 2005, \mnras, 364, L28

\bibitem[\protect\citeauthoryear{Rayner et~al.}{Rayner et~al.}{2003}]{Ray03}
Rayner, J.~T., Toomey, D.~W., Onaka, P.~M., Denault, A.~J., Stahlberger, W.~E.,
  Vacca, W.~D., Cushing, M.~C.,  \& Wang, S. 2003, PASP, 115, 362

\bibitem[\protect\citeauthoryear{{Rees}}{{Rees}}{1988}]{Rees88}
{Rees}, M.~J. 1988, \nat, 333, 523

\bibitem[\protect\citeauthoryear{Riffel, Rodr\'iguez-Ardila, \&
  Pastoriza}{Riffel et~al.}{2006}]{Rif06}
Riffel, R., Rodr\'iguez-Ardila, A.,  \& Pastoriza, M.~G. 2006, A\&A, 457, 61

\bibitem[\protect\citeauthoryear{{Rodr{\'{\i}}guez-Ardila}
  et~al.}{{Rodr{\'{\i}}guez-Ardila} et~al.}{2011}]{Rod11}
{Rodr{\'{\i}}guez-Ardila}, A., {Prieto}, M.~A., {Portilla}, J.~G.,  \&
  {Tejeiro}, J.~M. 2011, \apj, 743, 100

\bibitem[\protect\citeauthoryear{{Rodr{\'{\i}}guez-Ardila}
  et~al.}{{Rodr{\'{\i}}guez-Ardila} et~al.}{2002}]{Rod02c}
{Rodr{\'{\i}}guez-Ardila}, A., {Viegas}, S.~M., {Pastoriza}, M.~G.,  \&
  {Prato}, L. 2002, \apj, 579, 214

\bibitem[\protect\citeauthoryear{{Rose}, {Elvis}, \& {Tadhunter}}{{Rose}
  et~al.}{2015}]{Rose15}
{Rose}, M., {Elvis}, M.,  \& {Tadhunter}, C.~N. 2015, \mnras, 448, 2900

\bibitem[\protect\citeauthoryear{{Steenbrugge} et~al.}{{Steenbrugge}
  et~al.}{2005}]{KS05}
{Steenbrugge}, K.~C., et~al. 2005, \aap, 434, 569

\bibitem[\protect\citeauthoryear{{Tadhunter} \& {Tsvetanov}}{{Tadhunter} \&
  {Tsvetanov}}{1989}]{Tad89}
{Tadhunter}, C.,  \& {Tsvetanov}, Z. 1989, \nat, 341, 422

\bibitem[\protect\citeauthoryear{{Veilleux}}{{Veilleux}}{1988}]{Vei88}
{Veilleux}, S. 1988, \aj, 95, 1695

\bibitem[\protect\citeauthoryear{{Wang} et~al.}{{Wang} et~al.}{2012}]{WangT12}
{Wang}, T.-G., {Zhou}, H.-Y., {Komossa}, S., {Wang}, H.-Y., {Yuan}, W.,  \&
  {Yang}, C. 2012, \apj, 749, 115

\bibitem[\protect\citeauthoryear{{Wang} et~al.}{{Wang} et~al.}{2011}]{WangT11}
{Wang}, T.-G., {Zhou}, H.-Y., {Wang}, L.-F., {Lu}, H.-L.,  \& {Xu}, D. 2011,
  \apj, 740, 85

\bibitem[\protect\citeauthoryear{{Yang} et~al.}{{Yang} et~al.}{2013}]{Yang13}
{Yang}, C.-W., {Wang}, T.-G., {Ferland}, G., {Yuan}, W., {Zhou}, H.-Y.,  \&
  {Jiang}, P. 2013, \apj, 774, 46

\end{thebibliography}

\bsp
\label{lastpage}

\end{document}